\documentclass[prd,twocolumn,floatfix,nofootinbib,superscriptaddress]{revtex4}

\usepackage{graphicx}
\usepackage{bm}
\newcommand{\ligo}{{LIGO}}
\newcommand{\lisa}{{\sl LISA}}
\newcommand{\virgo}{{Virgo}}
\newcommand{\aigo}{{AIGO}}
\newcommand{\cross}{\times}
\newcommand{\vhat}[1]{\hat{\bm{#1}}}
\newcommand{\nhat}{\vhat{n}}
\newcommand{\Lhat}{\vhat{L}}

\begin{document}


\title{Short GRB and binary black hole standard sirens as a probe of dark
energy}

\author{Neal Dalal}
\affiliation{CITA, University of Toronto, 60 St. George St., Toronto,
  ON, M5S 3H8, Canada}

\author{Daniel E.\ Holz}
\affiliation{Theoretical Division, Los Alamos National Laboratory, Los Alamos, NM 87545}
\affiliation{Department of Astronomy and Astrophysics, University of Chicago, Chicago, IL 60637}

\author{Scott A.\ Hughes}
\affiliation{Dept.\ of Physics and MIT Kavli Institute, 77
  Massachusetts Avenue, Cambridge, MA 02139}

\author{Bhuvnesh Jain}
\affiliation{Dept.\ of Physics and Astronomy, University of
  Pennsylvania, Philadelphia, PA 19104}

\begin{abstract}
Observations of the gravitational radiation from well-localized,
inspiraling compact object binaries can measure absolute source
distances with high accuracy.  When coupled with an independent
determination of redshift through an electromagnetic counterpart,
these standard sirens can provide an excellent probe of the expansion
history of the Universe and the dark energy.  Short $\gamma$-ray
bursts, if produced by merging neutron star binaries, would be
standard sirens \textit{with known redshifts} detectable by
ground-based GW networks such as \ligo-II, \virgo, and \aigo.
Depending upon the collimation of these GRBs, a single year of
observation of their gravitational waves can measure the Hubble
constant $h$ to $\sim2\%$.  When combined with measurement of the
absolute distance to the last scattering surface of the cosmic
microwave background, this determines the dark energy equation of
state parameter $w$ to $\sim9\%$.  Similarly, supermassive binary
black hole inspirals will be standard sirens detectable by \lisa.
Depending upon the precise redshift distribution, $\sim 100$ sources
could measure $w$ at the $\sim4\%$ level.
\end{abstract}

\maketitle

\section{Introduction}

With the advent of the Laser Interferometer Gravitational-Wave
Observatory (\ligo), we are on the verge of an era of
gravitational-wave astronomy~\cite{barish99,ligo05}.
Among the most interesting expected sources for
GW observatories are the inspirals and mergers of compact-object
binaries.  LIGO-II, a planned upgrade with tenfold increase in
sensitivity, can detect
the inspirals and coalescence of stellar-mass binaries within
several hundred megaparsecs, while the Laser Interferometer Space
Antenna (\lisa) can study supermassive binary BHs ($M\sim10^4-10^7
M_\odot$) throughout the universe ($z\alt10$).

The idea of using GW measurements of coalescing binaries to make
cosmologically interesting measurements has a long history.  As
originally pointed out by \citet{schutz}, observation of the
gravitational radiation from an inspiraling binary provides a
self-calibrated absolute distance determination to the source.
\citet{chernoff93} and \citet{finn96} took advantage of this property
to show how, by observing many inspiral sources, one can construct the
distribution of observed binary mass and GW signal strength, and
thereby statistically constrain the values of cosmological parameters.
More recently, \citet{holz05} have shown that \lisa\ observations of
well-localized supermassive binary black hole (SMBBH) inspirals allow
cosmological distance determination with unprecedented accuracy, with
typical errors $<1\%$.  These GW ``standard sirens'' can precisely map
out the expansion history of the Universe, offering a powerful probe
of the dark energy.

The utility of standard sirens for constraining dark energy is quite
similar to that of standard candles, like Type-Ia supernovae.  One
advantage of GW standard sirens is that the underlying physics is
well-understood.  The radiation emitted during the inspiral phase (as
opposed to the merger phase) is well described using the
post-Newtonian expansion of general relativity for the BH binary
\cite{blanchet}.  Hence some unknown systematic evolution of the
standard sirens over time, mimicking a different cosmology, should not
be of concern.  Another advantage is that GW observatories directly
measure absolutely calibrated source distances, whereas Type-Ia
supernova standard candles provide only relatively calibrated
distances.

A major drawback of GW standard sirens is that, although the
gravitational waveforms measure distance directly,
they contain no redshift information. To be useful as a standard candle, an
independent measure of the redshift to the source is
crucial. This can be determined through observation of an
electromagnetic counterpart, such as the host galaxy of the
source. 
Unfortunately, as GW observatories are
essentially all-sky, they generally provide poor source
localization, and the host galaxy is not always
unambiguously identifiable \cite{zoltan05}.
In cases where source redshifts cannot be determined, the
distribution of unlocalized events can be used to place statistical
bounds on cosmological parameters \cite{finn96}.  However, in this
paper we will focus upon GW sirens whose redshifts may be measured, as
they can provide very tight constraints on cosmology.

Because standard siren distances are absolutely calibrated, even
sources at low redshift (e.g.\ $z\alt0.2$) can constrain dark energy.
This may seem surprising, since at low redshifts the
distance-redshift relation is well-described by a linear Hubble
relation $D = cz/H_0$, independent of dark energy parameters.  As
emphasized by \citet{hujain04} and \citet{hu04}, however, absolute
distances to 
sources at low redshift tightly constrain dark energy, when combined
with a determination of the absolute distance to the last-scattering
surface of the cosmic microwave background.  To understand this, note
that cosmological distances are given by a redshift integral of the
Hubble parameter, which in turn depends on the sum of energy
densities at each redshift:
\begin{eqnarray}
D(z_s)&=&\frac{c}{H_0\sqrt{\Omega_K}} \sinh\left[\sqrt{\Omega_K}
\int_0^{z_s}\frac{H_0}{H(z)}dz\right] \\
\frac{H(z)}{H_0}&=&
\sqrt{\Omega_m(1+z)^3+\Omega_{\rm de}(1+z)^{3(1+w)}+\Omega_K(1+z)^2}.
\nonumber
\end{eqnarray}
Here $\Omega_m+\Omega_{de}+\Omega_K=1$, $H_0 = 100\,h$
km/s/Mpc is the Hubble constant today, and we have assumed a
constant 
equation of state parameter $w$.  If we assume a flat universe
($\Omega_K=0$), then $\Omega_{\rm de}=1-\Omega_m$, and the only parameters
describing the global expansion are $h$, $\Omega_m$, and $w$.
Observations of the primary anisotropies in the cosmic microwave
background (CMB) provide two constraints on these three parameters.
First, the heights of the acoustic peaks determine the matter density
(in g/cm$^3$), which fixes $\Omega_m h^2$.  Second, the angular
scale of the peaks (their location in $l$-space) precisely measures
the angular diameter distance to the CMB last-scattering surface, in
Mpc.  Absolute distances to low-redshift sources measure
the Hubble constant $h$, which then allows all three parameters to be
determined \cite{hujain04,hu04}. Clearly the constraints we present
would be substantially degraded if the curvature were not fixed to
zero; see \cite{bernstein05,knox05} for the prospects for precise
constraints on curvature.

In addition to the low redshift standard sirens, those at higher redshifts also
help constrain dark energy, in the same manner as high-redshift standard
candles. \citet{holz05} discuss how \lisa\ observations of SMBBH
inspirals can help constrain cosmology. For a dark energy model which is
not dramatically different from a cosmological constant $\Lambda$, the
interesting redshift range is when the dark energy density is
significant ($z\alt1$), although note that gravitational lensing
degrades the constraints from the highest redshift standard sirens (or
candles) ~\cite{hl05}.

As mentioned above, the gravitational waves from standard sirens
measure source distances, but do not measure source
redshifts. Some sort
of electromagnetic counterpart associated with the merger event will
generally be required to use GW sources to determine cosmology.  
One potential class of GW sources guaranteed to have electromagnetic
counterparts are short $\gamma$-ray bursts (GRBs).  These sources are
thought to arise in the mergers of neutron star (NS) binaries, and hence
should be strong GW emitters in the frequency band accessible to
ground-based observatories. The GRB counterpart to these GW
source provides a precise sky localization, which is useful
both for determining the redshift to the source, and
for significantly improving the GW determination of absolute
distance.  As we discuss below, short GRBs occur at
a rate large enough for them to provide interesting
cosmological constraints.

\section{Distance determination for inspiraling binaries}

In this section we briefly review how distances to inspiraling
binaries may be determined; see Ref.~\cite{cf94} for more detail.  An
inspiraling binary at direction $\nhat$ on the sky, with orbital
angular momentum axis $\Lhat$, generates gravitational waves with
strain tensor
\begin{equation}
{\bf h}(t) = h_+(t) {\bf e}^+ + h_\cross(t) {\bf e}^\cross,
\end{equation}
where the basis tensors are
\begin{eqnarray}
{\bf e}^+ &=& \bm{e}_x\otimes\bm{e}_x - \bm{e}_y\otimes\bm{e}_y \\
{\bf e}^\cross &=& \bm{e}_x\otimes\bm{e}_y + \bm{e}_y\otimes\bm{e}_x
\end{eqnarray}
with 
\begin{eqnarray}
\bm{e}_x &=& \frac{\nhat\cross\Lhat}{|\nhat\cross\Lhat|}\\
\bm{e}_y &=& \bm{e}_x\cross\nhat.
\end{eqnarray}
Our convention is that $\nhat$ points towards the source, hence the
waves propagate in 
the direction $-\nhat$.  We can express the amplitudes of the two
polarizations $h_+(t)$ and $h_\cross(t)$ in the frequency domain as
\begin{equation}
{\tilde h}_+(f)=(1+v^2){\tilde h}_0(f), \quad 
{\tilde h}_\cross(f)=-2iv\,{\tilde h}_0(f) \label{pol}
\end{equation}
where $v\equiv\nhat\cdot\Lhat$ is the cosine of the inclination angle of the binary, and 
\begin{equation}
{\tilde h}_0(f)=\sqrt{\frac{5}{96}}\pi^{-2/3}
\left[\frac{G\cal M}{c^3}\right]^{5/6}\frac{c}{D}f^{-7/6}\exp[i\Psi(f)].
\end{equation}
In this expression, $D$ is the luminosity distance to the source, and
${\cal M}=(1+z)[m_1m_2]^{3/5}/(m_1+m_2)^{1/5}$ is the redshifted chirp
mass of the binary.  The phase $\Psi$ is given by 
\begin{equation}
\Psi(f)=2\pi f t_c - \phi_c - \frac{\pi}{4} + \frac{3}{4}
\left(\frac{8\pi G{\cal M} f}{c^3}\right)^{-5/3},
\end{equation} 
where $t_c$ is the time at coalescence, and $\phi_c$ is the orbital
phase at coalescence.

These expressions describe a binary's waves only in the Newtonian,
quadrupole approximation --- treating the binary's kinematics as due
to Newtonian gravity and using the quadrupole formula to estimate its
GW emission.  Because the phase parameters are essentially
uncorrelated from the amplitude parameters, this approximation is good
enough to estimate the expected signal-to-noise ratio from a source,
and provides a good estimate of the distance measurement accuracy, but
is not accurate enough to reliably model the detailed GW waveform
\cite{cf94}.  Higher order post-Newtonian templates (see Ref.\
\cite{blanchet} for detailed discussion) should be sufficiently
accurate, and are used for the actual data analysis.

Given ${\bf h}(t)$, the measured strain is given by 
\begin{equation}
h_M(t) = h^{ab}(t)d_{ab},
\end{equation}
where the detector response tensor for an interferometer with arms
$\vhat{l}$ and $\vhat{m}$ is ${\bf d}=(\vhat{l}\otimes\vhat{l} -
\vhat{m}\otimes\vhat{m})/2$.  In the notation of Ref.~\cite{cf94}, a
detector at colatitude $\theta$ and longitude $\phi$ with orientation
$\alpha$ has response tensor 
\begin{eqnarray}
{\bf d}&=&\cos(2\alpha)[\bm{e}_\theta\otimes\bm{e}_\phi +
\bm{e}_\phi\otimes\bm{e}_\theta]/2 \nonumber \\ && - \sin(2\alpha)
[\bm{e}_\theta\otimes\bm{e}_\theta - \bm{e}_\phi\otimes\bm{e}_\phi]/2.
\end{eqnarray}

To recap, the source parameters determining the measured signal are
distance $D$, chirp mass ${\cal M}$, coalescence time $t_c$,
coalescence phase $\phi_c$, source direction $\nhat$, and
orbital axis $\Lhat$.  These are the 8 parameters to be determined from
the data timestream $h_M(t)$.  If the detector has strain noise with
spectral density $S_h(f)$, then the incident strain is measured with
signal-to-noise ratio SNR (assuming Wiener filtering): 
\begin{equation}
{\rm SNR}^2 = 4 \int \frac{|\tilde h_M(f)|^2}{S_h(f)} df.
\end{equation}
The complicated angular dependence is hidden within the
measured strain $\tilde h_M$.  This dependence can be made more
explicit by rewriting the above equation as \cite{finn96}
\begin{equation}
\mbox{SNR}^2 = 4\frac{{\cal A}^2}{D^2} \left[F_+^2 (1 + v^2)^2 + 4
F_\times^2 v^2\right] I_7\;,
\label{eq:snr3}
\end{equation}
where ${\cal A} = \sqrt{5/96}\;\pi^{-2/3}\left(G{\cal M}/c^3\right)^{5/6}c$, 
$F_+={\bf e}^+_{ab}d^{ab}$, $F_\cross={\bf e}^\cross_{ab}d^{ab}$, and
\begin{equation}
I_7 = \int_{f_{\rm low}}^\infty \frac{f^{-7/3}}{S_h(f)}df\;.
\end{equation}
Here $f_{\rm low} \simeq 10$ Hz is the frequency below which the
detectors' sensitivities are badly degraded by ground motions.
In the optimal case, the binary is face-on ($v=1$) and directly
overhead, so that $F_+^2 + F_\cross^2 = 1$.  This gives 
\begin{equation}
\mbox{SNR}_{\rm opt} = 4\frac{{\cal A}}{D} I_7^{1/2}\;.
\label{eq:snr_opt}
\end{equation}
If instead we average over all sky positions and binary orientations,
we find 
\begin{equation}
\mbox{SNR}_{\rm ave} = \frac{8}{5}\frac{{\cal A}}{D} I_7^{1/2}\;,
\label{eq:snr_ave}
\end{equation}
where we have made use of $\langle F_+^2\rangle = \langle
F_\cross^2\rangle = 1/5$ and
\begin{eqnarray}
\frac{1}{2}\int_{-1}^1 (1+v^2)^2\,dv &=& \frac{28}{15}\;
\\
\frac{1}{2}\int_{-1}^1 4v^2\,dv &=& \frac{4}{3}\;.
\end{eqnarray}
Note that the SNR in the optimal geometry is a factor 5/2 times larger
than that for the average geometry.  Also note that face-on sources,
when averaged over all sky positions, have SNR a factor $\sqrt{5/4}
\simeq 1.12$ larger than SNR$_{\rm ave}$.

We can estimate how well the parameters $\bm{p}$ are measured using
the Fisher matrix 
\begin{equation}
F_{ij}=4 \int {\rm Re}\left[\frac{\partial_i \tilde h_M^\ast(f) \partial_j
    \tilde h_M(f)}{S_h(f)}\right] df,
\end{equation}
where $\partial_i\equiv\partial/\partial p_i$, and $^\ast$ denotes
complex conjugation.  Approximating the likelihood as
\begin{equation}
{\cal L} = \sqrt{\frac{|{\bf F}|}{(2\pi)^{n_p}}}\exp\left(-\frac{1}{2}\;
\Delta\bm{p}\cdot{\bf F}\cdot\Delta\bm{p}\right),
\end{equation}
then the error on parameter $p_i$ is given by
$\sqrt{\left({\bf F}^{-1}\right)_{ii}}$.  Prior constraints, or the
constraints from multiple detectors are implemented by multiplying the
respective likelihoods, which in this approximation reduces to summing
the respective Fisher matrices.  In our calculations, we compute
the partial derivatives numerically by finite differencing.

In practice, the `phase' parameters ${\cal M}$, $t_c$, and $\phi_c$
are determined with exquisite precision.  The `amplitude' parameters
$D$, $\Lhat$, and $\nhat$ are determined less well, in large part due
to parameter degeneracies.  By using multiple detectors many of these
degeneracies can be broken.  For example, timing information from a
network of detectors helps determine the source direction $\nhat$.
Similarly, if the detectors have different response tensors ${\bf d}$,
then the polarization of the GW signal may be measured, which
constrains the orbital axis $\Lhat$ (c.f.\ Eq.~(\ref{pol})).  

\section{GRBs observed by GW networks}

Short GRBs are an extremely promising source of gravitational waves
\cite{thorne87}.  These sources have been of great interest recently,
due to the prompt localization of the events by the
\textsl{Swift}\footnote{http://swift.gsfc.nasa.gov/docs/swift/swiftsc.html}\
\cite{gehrels05,berger05} and
\textsl{HETE-2}\footnote{http://space.mit.edu/HETE/Welcome.html}
\cite{fox05} satellites, allowing their detection in X-ray, optical,
and radio frequencies.  Particularly exciting has been the
identification of several galaxies hosting short bursts
\cite{gehrels05,fox05,shgrb}.  While the nature of short GRBs is not yet
known, a leading candidate is the merger of neutron star binaries
\cite{narayan92}, although other models have been proposed as well
\cite{macf05}.  The detection or non-detection of GRBs in
gravitational waves would of course be extremely useful \cite{finn99},
for example confirming or refuting the NS-NS merger scenario, or
determining the extent of collimation of the $\gamma$-ray emission
\cite{seto05}.

Additionally, as mentioned above, short GRBs can also be very useful
for determining the background cosmology by acting as GW standard
sirens.  One immediate advantage offered by GRBs is that their bright
electromagnetic emission allows a precise localization of the source
on the sky, pinpointing the source direction $\nhat$ and lifting some
of the degeneracies which limit distance determination.  The extent of
collimation in short GRBs is not well known, though indications of
beaming are claimed in at least one short GRB so far \cite{fox05}.  The
theoretical expectation is that emission should be beamed
preferentially along the orbital angular momentum axis where baryon
loading is minimized.  If this is the case, then we might expect short
GRBs to be nearly face-on, $v=\nhat\cdot\Lhat\approx 1$.  As can be
seen from Eq.~(\ref{pol}) this maximizes the amplitudes of both
polarizations and hence maximizes the SNR of the detection for a given
source direction $\nhat$.  In what follows, we will compute distance
errors for two cases: (1) isotropic distribution of $\Lhat$, and (2)
collimation, assuming an inclination probability distribution
$dP/dv\propto\exp(-(1-v)^2/2\sigma_v^2)$ for $\sigma_v=0.05$,
corresponding to a roughly $20^\circ$ jet angle.

The expected chirp mass for GRBs, ${\cal M}\approx 1.2 M_\odot$,
places them favorably in the frequency band accessible to ground-based
GW observatories. Several such observatories are now operating or are
planned for construction in the near future.  \ligo\ is already
operational, and its sensitivity should increase by an order of
magnitude in a planned upgrade (\ligo-II) \cite{ligonoise}.  A
detector of similar scale, \virgo\ \cite{virgo}, is under construction
in Italy, and there are plans for a similar detector, \aigo\
\cite{aigo}, in Australia.  The locations and orientations of these
observatories are listed in Table~\ref{coords}.  The two \ligo\
detectors are oriented to have very similar response tensors, and
therefore have limited ability to independently measure polarization
(and hence inclination).  Determining $\Lhat$ will thus require the
combination of \ligo\ with other observatories.

Henceforth, we assume that all four detectors will observe GRB events;
in subsequent work we intend to investigate how the distance errors
degrade if one or more elements of this network are removed.
Preliminary results indicate that reducing the size of the detector
network does not substantially degrade our ability to determine
distance (aside from the loss in total SNR) {\it assuming that we can
set a prior on the beaming factor} (and hence on the inclination
angle).  If we cannot set such a prior, then losing sites in this
network badly degrades our ability to determine distance to these
sources.  We emphasize this point to highlight the importance of
modeling bursts, and the importance of having widely separated GW
detectors around the globe.

\begin{table}[b]
\begin{ruledtabular}
\begin{tabular}{lccc}
Site & $\theta$ & $\phi$ & $\alpha$ \\
\hline
\ligo\ (Hanford) & 43.54 & -119.4 & 171 \\
\ligo\ (Livingston) & 59.44 & -90.77 & 243 \\
\virgo & 46.37 & 10.5 & 115.6 \\
\aigo & 121.4 & 115.7 & 45
\end{tabular}
\end{ruledtabular}
\caption{Coordinates of GW observatories, in the notation of
  Ref.~\protect{\cite{cf94}}.  All values are in degrees. \\
\label{coords}}
\end{table}

\begin{figure}
\centerline{\includegraphics[width=0.45\textwidth]{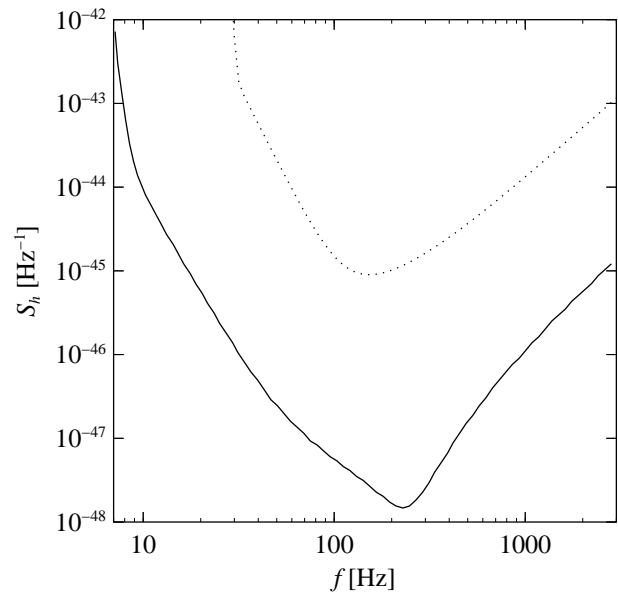}}
\caption{Noise curve for the \ligo\ detectors, for initial (dotted)
  and advanced (solid) sensitivity.\label{noise}}
\end{figure}

Figure~\ref{noise} plots the noise spectral density forecasted for
\ligo-II \cite{ligonoise}.  Projected noise curves for the advanced
detector configurations are not yet available for \virgo\ or \aigo, so
for simplicity we use the \ligo-II curve for all the observatories in
the network.  For comparison, we also show the sensitivity for the
currently operating LIGO observatories.

With the response tensors for the elements in our network, and their
noise spectra, we can now compute the Fisher matrices and parameter
errors for GRBs as a function of distance and location on the sky.
For convenience, when computing the Fisher matrix we replace the
parameter pair $\{D,v=\nhat\cdot\Lhat\}$ with the pair $\{(1+v)^2/D,
(1-v)^2/D\}$, to avoid singularities in the limit $v\to 1$ when $v$
and $D$ become degenerate \cite{cf94}.  Another difficulty that arises
in the face-on limit is that the position angle of $\Lhat$,
denoted $\psi$ by Ref.~\cite{cf94}, becomes meaningless as $v\to1$.
Including it as a parameter would cause the Fisher matrix to become
singular in the face-on limit; we circumvent this difficulty using
singular value decomposition to invert the Fisher matrix, zeroing
any eigenvalue whose magnitude is $10^{-10}$ times that of the largest
eigenvalue.  

\begin{figure}
\centerline{\includegraphics[width=0.4\textwidth]{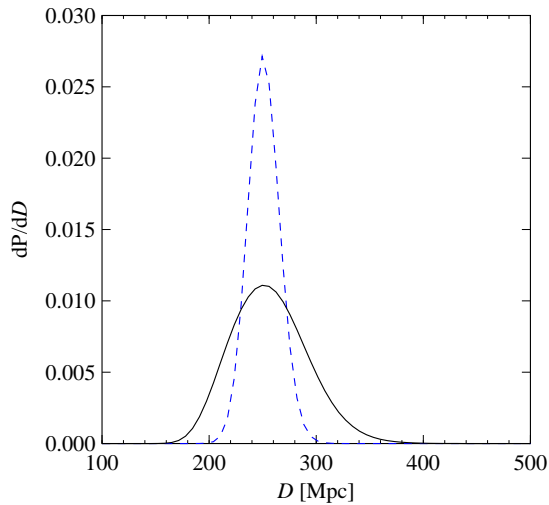}}
\caption{Distribution of measured distances for a source at $D=250$
  Mpc, averaged over 100 source directions $\nhat$ and orientations
  $\Lhat$. The solid curve shows constraints for randomly oriented
  sources, while the dashed curve shows constraints for collimated
  sources with $\sigma_v=0.05$.  \label{d250}}
\end{figure}

Because the antenna response of each detector varies strongly with
source direction $\nhat$, the parameter errors at any given distance
$D$ also depend strongly on $\nhat$.  We are interested only in
average errors as a function of $D$; hence, for each $D$ we average
over 100 different orientations $\Lhat$ and $\nhat$.  For example,
Fig.~\ref{d250} shows the expected constraints for sources at distance
$D=250$ Mpc.  Note that the errors significantly improve if it is
assumed that sources are beamed towards us.

\begin{figure}
\centerline{\includegraphics[width=0.4\textwidth]{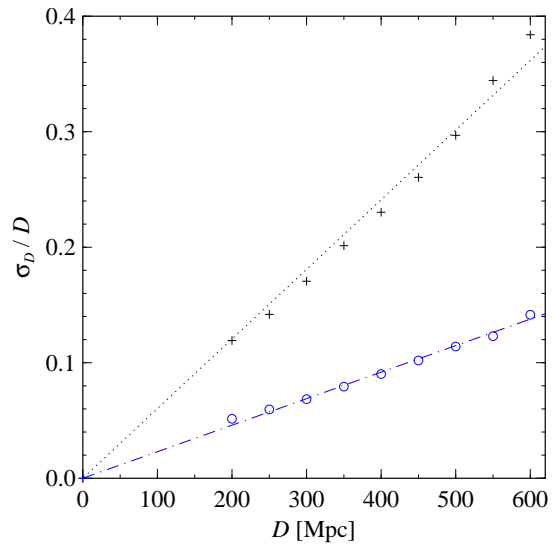}}
\caption{Fractional distance errors as a function of source distance
  $D$.  The + symbols are for unbeamed GRBs, while circles are for
  $\sigma_v=0.05$.  The two lines show the best-fit linear relations;
  note that there may be departures from linear scaling at the highest
  distances.
  \label{d_errs}}
\end{figure}

Given the likelihood distribution $dP/dD$, we define the distance
error as $\sigma_D^2 = \langle D^2\rangle - \langle D\rangle^2$, where
averages are with respect to $dP/dD$.  Figure~\ref{d_errs} plots
$\sigma_D$ as a function of $D$.  Our results appear roughly
consistent with $\sigma_D/D\propto D \propto 1/\mbox{SNR}$.  Our
best-fit linear scaling for unbeamed GRBs is 
$\sigma_D/D=D/(1.7\mbox{ Gpc})$, and $\sigma_D/D=D/(4.4\mbox{ Gpc})$
for collimation $\sigma_v=0.05$.  Henceforth, we will assume these
scalings when estimating cosmological constraints from GW network
observations of short GRBs.

\section{cosmological constraints from standard sirens}

As discussed in \S 1, a measurement of the Hubble constant $h$ using
GRBs, when combined with CMB constraints, also constrains dark energy
parameters. 
We use two measurements from the CMB: determination of
the angular scale of the acoustic peaks $l_A$, and determination of
the matter density $\Omega_m h^2$ from the peak heights.  Currently,
the \textsl{WMAP} satellite has measured $l_A=300\pm3$ and $\Omega_m
h^2=0.14 \pm 0.02$ \cite{page03}.  The error on $\Omega_m h^2$ will
soon decrease by a factor $\sim\sqrt{3}$ with the 3rd year release of
\textsl{WMAP} data.  The \textsl{Planck} satellite is expected to measure
$\Omega_m h^2$ to a fractional error of $\sim 1\%$. 

The acoustic scale $l_A = \pi D_\star/s_\star$, where $D_\star$ is the
distance to the last-scattering surface at $z=1089$, and $s_\star$ is
the sound horizon at decoupling, approximately given by $s_\star=144.4
\mbox{ Mpc } (\Omega_m h^2/0.14)^{-0.252}$ \cite{hu04}.  Given the
dependence of these observables on the cosmological parameters
$\bm{p}=\{h,\Omega_m,w\}$, we can then estimate parameter errors
using the Fisher matrix:
\begin{eqnarray}
F_{ij}&=&\frac{\partial_i l_A \partial_j l_A}{\sigma_A^2}
+ \frac{\partial_i \Omega_m h^2 \partial_j \Omega_m
  h^2}{\sigma_{\omega_m}^2}  \label{fish}\nonumber\\
&& + \int_0^{z_{\rm max}}\frac{dN}{dz}\frac{\partial_iD_L(z)
\partial_jD_L(z)}{\sigma_D(z)^2 + \left(\sigma_z\frac{dD_L}{dz}\right)^2} dz ,
\end{eqnarray}
where redshift errors $\sigma_z$ are caused by peculiar 
velocities\footnote{It may be preferable to measure redshifts of the
  host galaxies rather than the GRBs themselves, whose progenitors may
  suffer kicks which will add in quadrature to the redshift noise from
  peculiar velocities.}
with assumed rms of 300 km/s.  The luminosity distance $D_L(z)=(1+z)D(z)$,
and its error $\sigma_D$ includes both GW errors, as computed
in the previous section, and gravitational lensing errors
\cite{dalal03}, computed using an approximate nonlinear power spectrum
\cite{smith03}.  

For the source redshift distribution $dN/dz$, we
assume that short GRBs occur at a constant comoving rate of 10
Gpc$^{-3}$ yr$^{-1}$ \cite{nakar05}.  We found in the previous section
that the SNR in distance determination per source scales roughly like
$1/D$.  Since the number 
of sources scales with volume $\propto D^3$, we expect the SNR on the
Hubble constant $h$ to scale like $D_{\rm max}^{1/2}$, where 
$D_{\rm max}$ is the maximum distance to which GRBs may be detected as
gravitational wave sources. 

The standard threshold used in the GW literature for detection has
been SNR $>8.5$ \cite{cf94,flanagan98}.  The reason for this high
threshold is that sources are detected by correlating the data
timestream with large numbers (e.g. $10^{15}$) of templates
corresponding to different parameter values, and therefore the
detection threshold must be set high to avoid excessive numbers of false
detections.  Such large numbers of templates are required in order to
fully explore parameter space.  For GRB sources, however, the
parameter space to be searched is considerably reduced: the
$\gamma$-ray burst itself determines the source direction $\nhat$ and
time $t_c$.  Depending upon one's confidence in theoretical models for
GRBs, the chirp mass ${\cal M}$ and orientation $\Lhat$ may also
constrained.  Because many fewer templates need to be run for GRB
sources, we should set the detection threshold correspondingly lower.
We conservatively estimate that knowledge of the GRB time
reduces the number of required templates by a factor $\sim
10^5$, corresponding to a reduced threshold SNR $>7$.  Note that
this is the \textit{total} SNR; since we have assumed a network of
four detectors with identical noise, this translates into a threshold
SNR $>3.5$ per detector.  From this, we can determine the maximum
distance to which sources may be detected using
Eq.~(\ref{eq:snr_ave}).  For chirp mass ${\cal M}=1.2 M_\odot$, we
have ${\cal A}=4.7\times10^{-6} s^{5/6}$, and for our assumed noise
spectral density (Fig.~\ref{noise}), 
$I_7=8.33\times 10^{44}{\rm Hz}^{-1/3}$.  Therefore the maximum
distance for which SNR$_{\rm ave}>3.5$ is $D_{\rm max}=600$ Mpc.  

\begin{figure}
\centerline{\includegraphics[width=0.4\textwidth]{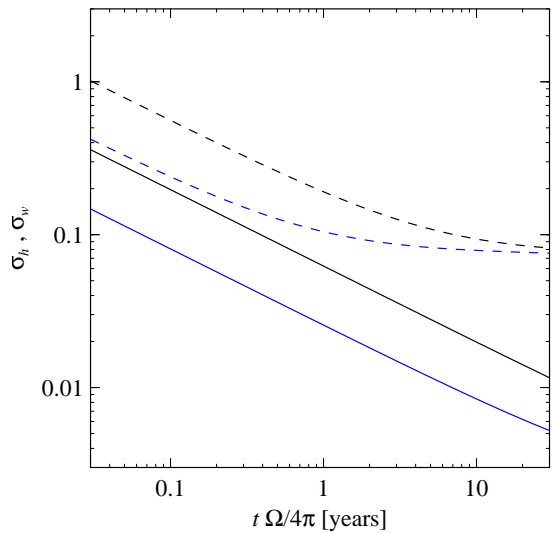}}
\caption{Errors on $h$ and $w$ as a function of GRB exposure, assuming
  \textsl{Planck}-quality errors from CMB. Solid
  curves are for $\sigma_h$, the error on the Hubble constant, while
  dashed curves correspond to $\sigma_w$, for the dark energy equation
  of state parameter.  The lower curves are for beamed GRBs with
  $\sigma_v=0.05$ while the upper curves are for unbeamed GRBs.
\label{sigt}}
\end{figure}

Assuming default cosmological parameters $h=0.72$, $\Omega_m=0.27$, and
$w=-1$, the resulting parameter errors computed from Eq.~(\ref{fish})
are shown in Figure~\ref{sigt}, as a function of the time and sky area
over which GRBs are observed.  While errors on the Hubble constant
scale like $\sigma_h\propto N_{\rm GRB}^{-1/2}$, the errors on $w$
scale this way only in the limit of small numbers of sources.  Quite
rapidly, the limiting error on $w$ becomes the uncertainty in CMB (in
this figure, fractional errors of 1\% on $\Omega_m h^2$ were
assumed).  Unless CMB errors can be significantly improved,
it will be difficult for low-redshift GW sources to
constrain $w$ to better than the $\sim10\%$ level.

Higher redshift standard sirens would probe departures of the cosmic
expansion from linear Hubble scaling, and thereby directly constrain
parameters like $\Omega_m$ and $w$. Unfortunately, stellar-mass
inspirals at high redshift
are not sufficiently luminous to be detected by any existing
or planned GW observatory.  Inspirals involving supermassive black
hole binaries, however, are sufficiently luminous in gravitational
waves to be detected at cosmological distances.  As discussed by
\citet{holz05}, \lisa\ observations of SMBBH inspirals can in principle
measure distances to better than 1\% accuracy.  This precision is
degraded, however, by gravitational lensing caused by density
fluctuations from large-scale structure along the line of sight to the
source.  Another difficulty in using \lisa\ observations is that, unlike in the
case of short GRBs, for SMBBHs there are no guaranteed electromagnetic
counterparts.  However, it has been argued that many SMBBH 
mergers will be followed by bright quasar-like activity
\cite{milos05}, or possibly preceded by optical emission \cite{armitage02},
which will localize the GW source on the sky and
provide a source redshift.

\begin{figure}
\centerline{\includegraphics[width=0.45\textwidth]{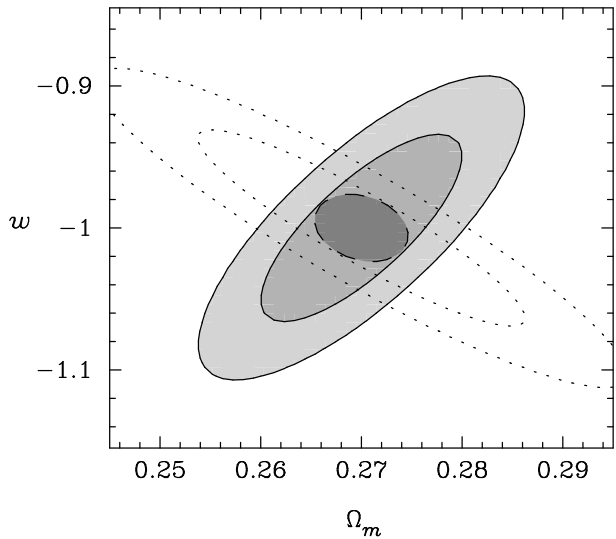}}
\caption{\lisa\ constraints on dark energy.  The solid contours show
  the 68\% and 95\% confidence regions expected for a sample of 100
  SMBBH sources observed by \lisa, distributed with constant comoving
  density between $0<z<2$.  A \textsl{Planck}\ prior also has been
  used on $\Omega_m h^2$ and $l_A$, as discussed in the text.  The
  dotted contours correspond to a sample of 3000 SNe with intrinsic
  luminosity scatter of 10\%, with redshift distribution $\propto
  \exp(-(z-0.5)^2)$ over $0.02<z<2$.  The dashed (dark shaded) contour
  shows the 68\% confidence region for the combined constraints
  GW+SNe+CMB.
\label{con}}
\end{figure}

Due to lensing errors, small numbers of \lisa\ sources will generally be unable
to constrain dark energy parameters significantly \cite{holz05}.
The effects of lensing diminish significantly at lower redshifts,
so a single SMBBH inspiral at $z<0.5$ observed by \lisa\ could
measure the Hubble constant to $\alt1\%$ and $w$ to $\alt10\%$.
Although such a source is unlikely,
the low redshift regime should already be well-determined by
ground-based GW observations of short GRBs.  On the other hand, if
large numbers of SMBBH mergers occur during \lisa's lifetime, then
\lisa\ should provide quite interesting constraints on dark energy,
despite the lensing noise.  To illustrate this, Figure~\ref{con} plots expected
constraints in the $\Omega_m$ vs.\ $h$ plane for a sample of 100 SMBBH
inspirals observed by \lisa, distributed in redshift assuming a
constant comoving density between $0<z<2$, combined with constraints
from \textsl{Planck}-quality CMB data. The $1-\sigma$ errors on
$w$ are $\sigma_w=0.04$; these are competitive with ambitious Type-Ia
supernova surveys like JDEM.  Note that these errors improve
considerably if the main source of noise, gravitational lensing, can
be cleaned out by reconstructing the lensing mass distribution using
other probes.  \citet{dalal03} argue that cosmic shear measured from
optical surveys would not allow mass reconstruction with sufficient
angular resolution.  Cosmic magnification measured in the radio could
conceivably offer an alternative route (e.g. \cite{pen04}).

Our discussion has focused on gravitational lensing only as a source of
noise, but in principle there is cosmological information which
can be extracted from the lensing fluctuations themselves.  With large
numbers of sources, \lisa\ observations of cosmic magnification can
provide constraints complementary to other probes.  We
would not expect GW standard sirens to usefully probe the power
spectra of matter fluctuations or galaxy-mass correlations
\cite{hujain04} at any scale, compared to other means like cosmic
shear or Type-Ia supernovae, based on their noise power spectra :
\begin{equation}
\frac{\gamma_{\rm gal}^2}{n_{\rm gal}} \ll 
\frac{\sigma_{\rm SN}^2}{n_{\rm SN}} \ll 
\frac{\sigma_{\rm GW}^2}{n_{\rm GW}},
\end{equation}
where galaxies have shape noise $\gamma_{\rm gal}\approx 0.4$ and
number density $n_{\rm gal}\approx 50/{\rm arcmin}^2$, supernovae have
luminosity dispersion $\sigma_{\rm SN}\approx 0.1$ and number density
$n_{\rm SN}\approx 4000/(20\ {\rm deg})^2$ as observed by SNAP, and GW
standard sirens have luminosity errors $\sigma_{\rm GW}\approx 1\%$
and number density $n_{\rm GW}\approx 100/(4\pi\ {\rm sr})$.  On the
other hand, GW standard sirens can determine 1-point functions of the
matter density better than other methods, in particular the
probability distribution of lensing magnification.  This could be
useful for distinguishing between different dark matter models
\cite{seljakholz}.  

\section{Discussion}

We have shown that observations of the gravitational waves emitted by
binary compact object inspirals can be a powerful probe of cosmology.
In particular, short $\gamma$-ray bursts appear quite promising as
potential GW standard sirens.  The presently observed rate of short
GRBs is sufficiently high that within a few years of observation by
the next generation of ground-based GW observatories (e.g. \ligo-II,
\virgo\ and \aigo), strong constraints on dark energy parameters may
be derived ($\sigma_w<0.1$).  These inspiraling NS-binaries should be
clean sources of gravitational waves; possible sources of
contamination, such as tidal effects, magnetic torques, or
gasdynamical torques from circumbinary gas, should all be negligible
during the crucial inspiral phase (where $v/c \alt 0.3$).  We
emphasize again that the best information about distance measurements
comes from combining multiple GW data from instruments that are widely
separated.  Good information about the collimation of the gamma rays
and thus on the likely inclination of the binary progenitor will also
improve the utility of these standard sirens.  Given the great
cosmological potential of GW observations of short GRBs, there is
strong incentive to extend the lifetime of GRB satellites such as
\textsl{Swift} or \textsl{HETE}-2 to overlap with next-generation
gravitational wave observatories.

The inspirals of SMBBH binaries observed by \lisa\ can also provide
interesting constraints on dark energy, if the rate of such mergers is
high enough to average away noise caused by gravitational lensing.  At
present, the total rate and redshift distribution of SMBBH mergers are
not well understood, with estimates ranging from a few (or zero) per
year, up to hundreds per year, depending upon assumptions
\cite{sesana04,menou01,haehnelt03,koushiappas06}. If the rates are at
the high end of these estimates, with a significant fraction at
redshifts $z<2$, then $w$ may be constrained at the few percent level.

\begin{acknowledgments}
We thank Olivier Dor\'e, {\'E}anna Flanagan, Wendy Freedman, Samaya
Nissanke, Mike Nolta, Maria Papa, 
Sterl Phinney, Roman Rafikov, and Ravi Sheth for
useful discussions.  David Blair, Raffaele Flaminio, and David
Shoemaker provided updated coordinates and orientations for AIGO, LIGO
and Virgo.  We also thank Martin Hendry and the organizers of the ETSU
workshop on gravitational waves, where this work was initiated.  ND is
supported by CITA and the National Science and Engineering Research
Council of Canada. DEH acknowledges a Feynman Fellowship from LANL, 
and financial support from Willie Nelson.
SAH is supported by NSF grants PHY-0244424 and PHY-0449884, by NASA
Grants NAG5-12906 and NNG05G105G, and by MIT's Class of 1956 Career
Development Fund.  BJ is supported by NASA grant NAG5-10924 and by 
NSF grant AST03-07297. 
\end{acknowledgments}

\newcommand{\apjl}{\apj\ Letters}
\newcommand{\apjs}{\apj\ Supp.}
\newcommand{\mnras}{Mon.\ Not. Royal Astron. Soc.}
\newcommand{\aap}{Astron.\ \& Astrophys.}


\begin{thebibliography}{39}
\expandafter\ifx\csname natexlab\endcsname\relax\def\natexlab#1{#1}\fi
\expandafter\ifx\csname bibnamefont\endcsname\relax
  \def\bibnamefont#1{#1}\fi
\expandafter\ifx\csname bibfnamefont\endcsname\relax
  \def\bibfnamefont#1{#1}\fi
\expandafter\ifx\csname citenamefont\endcsname\relax
  \def\citenamefont#1{#1}\fi
\expandafter\ifx\csname url\endcsname\relax
  \def\url#1{\texttt{#1}}\fi
\expandafter\ifx\csname urlprefix\endcsname\relax\def\urlprefix{URL }\fi
\providecommand{\bibinfo}[2]{#2}
\providecommand{\eprint}[2][]{\url{#2}}

\bibitem[{\citenamefont{Abbott et~al.}(2005)}]{ligo05}
\bibinfo{author}{\bibfnamefont{B.}~\bibnamefont{Abbott}} \bibnamefont{et~al.},
  \bibinfo{journal}{\prd} \textbf{\bibinfo{volume}{72}},
  \bibinfo{pages}{102004} (\bibinfo{year}{2005}).

\bibitem[{\citenamefont{Barish and Weiss}(1999)}]{barish99}
\bibinfo{author}{\bibfnamefont{B.~C.} \bibnamefont{Barish}} \bibnamefont{and}
  \bibinfo{author}{\bibfnamefont{R.}~\bibnamefont{Weiss}},
  \bibinfo{journal}{Phys. Today} \textbf{\bibinfo{volume}{52N10}},
  \bibinfo{pages}{44} (\bibinfo{year}{1999}).

\bibitem[{\citenamefont{{Schutz}}(1986)}]{schutz}
\bibinfo{author}{\bibfnamefont{B.~F.} \bibnamefont{{Schutz}}},
  \bibinfo{journal}{\nat} \textbf{\bibinfo{volume}{323}}, \bibinfo{pages}{310}
  (\bibinfo{year}{1986}).

\bibitem[{\citenamefont{{Chernoff} and {Finn}}(1993)}]{chernoff93}
\bibinfo{author}{\bibfnamefont{D.~F.} \bibnamefont{{Chernoff}}}
  \bibnamefont{and} \bibinfo{author}{\bibfnamefont{L.~S.}
  \bibnamefont{{Finn}}}, \bibinfo{journal}{\apjl}
  \textbf{\bibinfo{volume}{411}}, \bibinfo{pages}{L5} (\bibinfo{year}{1993}).

\bibitem[{\citenamefont{{Finn}}(1996)}]{finn96}
\bibinfo{author}{\bibfnamefont{L.~S.} \bibnamefont{{Finn}}},
  \bibinfo{journal}{\prd} \textbf{\bibinfo{volume}{53}}, \bibinfo{pages}{2878}
  (\bibinfo{year}{1996}).

\bibitem[{\citenamefont{{Holz} and {Hughes}}(2005)}]{holz05}
\bibinfo{author}{\bibfnamefont{D.~E.} \bibnamefont{{Holz}}} \bibnamefont{and}
  \bibinfo{author}{\bibfnamefont{S.~A.} \bibnamefont{{Hughes}}},
  \bibinfo{journal}{\apj} \textbf{\bibinfo{volume}{629}}, \bibinfo{pages}{15}
  (\bibinfo{year}{2005}).

\bibitem[{\citenamefont{{Blanchet}}(2002)}]{blanchet}
\bibinfo{author}{\bibfnamefont{L.}~\bibnamefont{{Blanchet}}},
  \bibinfo{journal}{Living Reviews in Relativity} \textbf{\bibinfo{volume}{5}},
  \bibinfo{pages}{3} (\bibinfo{year}{2002}).

\bibitem[{\citenamefont{{Kocsis} et~al.}(2005)\citenamefont{{Kocsis}, {Frei},
  {Haiman}, and {Menou}}}]{zoltan05}
\bibinfo{author}{\bibfnamefont{B.}~\bibnamefont{{Kocsis}}},
  \bibinfo{author}{\bibfnamefont{Z.}~\bibnamefont{{Frei}}},
  \bibinfo{author}{\bibfnamefont{Z.}~\bibnamefont{{Haiman}}}, \bibnamefont{and}
  \bibinfo{author}{\bibfnamefont{K.}~\bibnamefont{{Menou}}},
  \bibinfo{journal}{ArXiv Astrophysics e-prints}  (\bibinfo{year}{2005}),
  \eprint{arXiv:astro-ph/0505394}.

\bibitem[{\citenamefont{{Hu} and {Jain}}(2004)}]{hujain04}
\bibinfo{author}{\bibfnamefont{W.}~\bibnamefont{{Hu}}} \bibnamefont{and}
  \bibinfo{author}{\bibfnamefont{B.}~\bibnamefont{{Jain}}},
  \bibinfo{journal}{\prd} \textbf{\bibinfo{volume}{70}},
  \bibinfo{pages}{043009} (\bibinfo{year}{2004}).

\bibitem[{\citenamefont{{Hu}}(2004)}]{hu04}
\bibinfo{author}{\bibfnamefont{W.}~\bibnamefont{{Hu}}},
  \bibinfo{journal}{astro-ph/0407158}  (\bibinfo{year}{2004}).

\bibitem[{\citenamefont{{Bernstein}}(2005)}]{bernstein05}
\bibinfo{author}{\bibfnamefont{G.}~\bibnamefont{{Bernstein}}},
  \bibinfo{journal}{ArXiv Astrophysics e-prints}  (\bibinfo{year}{2005}),
  \eprint{arXiv:astro-ph/0503276}.

\bibitem[{\citenamefont{{Knox}}(2005)}]{knox05}
\bibinfo{author}{\bibfnamefont{L.}~\bibnamefont{{Knox}}},
  \bibinfo{journal}{ArXiv Astrophysics e-prints}  (\bibinfo{year}{2005}),
  \eprint{arXiv:astro-ph/0503405}.

\bibitem[{\citenamefont{{Holz} and {Linder}}(2005)}]{hl05}
\bibinfo{author}{\bibfnamefont{D.~E.} \bibnamefont{{Holz}}} \bibnamefont{and}
  \bibinfo{author}{\bibfnamefont{E.~V.} \bibnamefont{{Linder}}},
  \bibinfo{journal}{\apj} \textbf{\bibinfo{volume}{631}}, \bibinfo{pages}{678}
  (\bibinfo{year}{2005}).

\bibitem[{\citenamefont{{Cutler} and {Flanagan}}(1994)}]{cf94}
\bibinfo{author}{\bibfnamefont{C.}~\bibnamefont{{Cutler}}} \bibnamefont{and}
  \bibinfo{author}{\bibfnamefont{{\'E}.~E.} \bibnamefont{{Flanagan}}},
  \bibinfo{journal}{\prd} \textbf{\bibinfo{volume}{49}}, \bibinfo{pages}{2658}
  (\bibinfo{year}{1994}).

\bibitem[{\citenamefont{Thorne}(1987)}]{thorne87}
\bibinfo{author}{\bibfnamefont{K.~S.} \bibnamefont{Thorne}}
  (\bibinfo{year}{1987}), \bibinfo{note}{in *Hawking, S.W. (ed.), Israel, W.
  (ed.): Three hundred years of gravitation*, 330-458. (see Book Index)}.

\bibitem[{\citenamefont{{Gehrels} et~al.}(2005)\citenamefont{{Gehrels},
  {Sarazin}, {O'Brien}, {Zhang}, {Barbier}, {Barthelmy}, {Blustin}, {Burrows},
  {Cannizzo}, {Cummings} et~al.}}]{gehrels05}
\bibinfo{author}{\bibfnamefont{N.}~\bibnamefont{{Gehrels}}},
  \bibinfo{author}{\bibfnamefont{C.~L.} \bibnamefont{{Sarazin}}},
  \bibinfo{author}{\bibfnamefont{P.~T.} \bibnamefont{{O'Brien}}},
  \bibinfo{author}{\bibfnamefont{B.}~\bibnamefont{{Zhang}}},
  \bibinfo{author}{\bibfnamefont{L.}~\bibnamefont{{Barbier}}},
  \bibinfo{author}{\bibfnamefont{S.~D.} \bibnamefont{{Barthelmy}}},
  \bibinfo{author}{\bibfnamefont{A.}~\bibnamefont{{Blustin}}},
  \bibinfo{author}{\bibfnamefont{D.~N.} \bibnamefont{{Burrows}}},
  \bibinfo{author}{\bibfnamefont{J.}~\bibnamefont{{Cannizzo}}},
  \bibinfo{author}{\bibfnamefont{J.~R.} \bibnamefont{{Cummings}}},
  \bibnamefont{et~al.}, \bibinfo{journal}{\nat} \textbf{\bibinfo{volume}{437}},
  \bibinfo{pages}{851} (\bibinfo{year}{2005}).

\bibitem[{\citenamefont{{Berger} et~al.}(2005)\citenamefont{{Berger}, {Price},
  {Cenko}, {Gal-Yam}, {Soderberg}, {Kasliwal}, {Leonard}, {Cameron}, {Frail},
  {Kulkarni} et~al.}}]{berger05}
\bibinfo{author}{\bibfnamefont{E.}~\bibnamefont{{Berger}}},
  \bibinfo{author}{\bibfnamefont{P.~A.} \bibnamefont{{Price}}},
  \bibinfo{author}{\bibfnamefont{S.~B.} \bibnamefont{{Cenko}}},
  \bibinfo{author}{\bibfnamefont{A.}~\bibnamefont{{Gal-Yam}}},
  \bibinfo{author}{\bibfnamefont{A.~M.} \bibnamefont{{Soderberg}}},
  \bibinfo{author}{\bibfnamefont{M.}~\bibnamefont{{Kasliwal}}},
  \bibinfo{author}{\bibfnamefont{D.~C.} \bibnamefont{{Leonard}}},
  \bibinfo{author}{\bibfnamefont{P.~B.} \bibnamefont{{Cameron}}},
  \bibinfo{author}{\bibfnamefont{D.~A.} \bibnamefont{{Frail}}},
  \bibinfo{author}{\bibfnamefont{S.~R.} \bibnamefont{{Kulkarni}}},
  \bibnamefont{et~al.}, \bibinfo{journal}{\nat} \textbf{\bibinfo{volume}{438}},
  \bibinfo{pages}{988} (\bibinfo{year}{2005}).

\bibitem[{\citenamefont{{Fox} et~al.}(2005)\citenamefont{{Fox}, {Frail},
  {Price}, {Kulkarni}, {Berger}, {Piran}, {Soderberg}, {Cenko}, {Cameron},
  {Gal-Yam} et~al.}}]{fox05}
\bibinfo{author}{\bibfnamefont{D.~B.} \bibnamefont{{Fox}}},
  \bibinfo{author}{\bibfnamefont{D.~A.} \bibnamefont{{Frail}}},
  \bibinfo{author}{\bibfnamefont{P.~A.} \bibnamefont{{Price}}},
  \bibinfo{author}{\bibfnamefont{S.~R.} \bibnamefont{{Kulkarni}}},
  \bibinfo{author}{\bibfnamefont{E.}~\bibnamefont{{Berger}}},
  \bibinfo{author}{\bibfnamefont{T.}~\bibnamefont{{Piran}}},
  \bibinfo{author}{\bibfnamefont{A.~M.} \bibnamefont{{Soderberg}}},
  \bibinfo{author}{\bibfnamefont{S.~B.} \bibnamefont{{Cenko}}},
  \bibinfo{author}{\bibfnamefont{P.~B.} \bibnamefont{{Cameron}}},
  \bibinfo{author}{\bibfnamefont{A.}~\bibnamefont{{Gal-Yam}}},
  \bibnamefont{et~al.}, \bibinfo{journal}{\nat} \textbf{\bibinfo{volume}{437}},
  \bibinfo{pages}{845} (\bibinfo{year}{2005}).

\bibitem[{\citenamefont{{Barthelmy} et~al.}(2005)\citenamefont{{Barthelmy},
  {Chincarini}, {Burrows}, {Gehrels}, {Covino}, {Moretti}, {Romano}, {O'Brien},
  {Sarazin}, {Kouveliotou} et~al.}}]{shgrb}
\bibinfo{author}{\bibfnamefont{S.~D.} \bibnamefont{{Barthelmy}}},
  \bibinfo{author}{\bibfnamefont{G.}~\bibnamefont{{Chincarini}}},
  \bibinfo{author}{\bibfnamefont{D.~N.} \bibnamefont{{Burrows}}},
  \bibinfo{author}{\bibfnamefont{N.}~\bibnamefont{{Gehrels}}},
  \bibinfo{author}{\bibfnamefont{S.}~\bibnamefont{{Covino}}},
  \bibinfo{author}{\bibfnamefont{A.}~\bibnamefont{{Moretti}}},
  \bibinfo{author}{\bibfnamefont{P.}~\bibnamefont{{Romano}}},
  \bibinfo{author}{\bibfnamefont{P.~T.} \bibnamefont{{O'Brien}}},
  \bibinfo{author}{\bibfnamefont{C.~L.} \bibnamefont{{Sarazin}}},
  \bibinfo{author}{\bibfnamefont{C.}~\bibnamefont{{Kouveliotou}}},
  \bibnamefont{et~al.}, \bibinfo{journal}{ArXiv Astrophysics e-prints}
  (\bibinfo{year}{2005}), \eprint{arXiv:astro-ph/0511579}.

\bibitem[{\citenamefont{{Narayan} et~al.}(1992)\citenamefont{{Narayan},
  {Paczynski}, and {Piran}}}]{narayan92}
\bibinfo{author}{\bibfnamefont{R.}~\bibnamefont{{Narayan}}},
  \bibinfo{author}{\bibfnamefont{B.}~\bibnamefont{{Paczynski}}},
  \bibnamefont{and} \bibinfo{author}{\bibfnamefont{T.}~\bibnamefont{{Piran}}},
  \bibinfo{journal}{\apjl} \textbf{\bibinfo{volume}{395}}, \bibinfo{pages}{L83}
  (\bibinfo{year}{1992}).

\bibitem[{\citenamefont{{MacFadyen} et~al.}(2005)\citenamefont{{MacFadyen},
  {Ramirez-Ruiz}, and {Zhang}}}]{macf05}
\bibinfo{author}{\bibfnamefont{A.~I.} \bibnamefont{{MacFadyen}}},
  \bibinfo{author}{\bibfnamefont{E.}~\bibnamefont{{Ramirez-Ruiz}}},
  \bibnamefont{and} \bibinfo{author}{\bibfnamefont{W.}~\bibnamefont{{Zhang}}},
  \bibinfo{journal}{ArXiv Astrophysics e-prints}  (\bibinfo{year}{2005}),
  \eprint{arXiv:astro-ph/0510192}.

\bibitem[{\citenamefont{{Finn} et~al.}(1999)\citenamefont{{Finn}, {Mohanty},
  and {Romano}}}]{finn99}
\bibinfo{author}{\bibfnamefont{L.~S.} \bibnamefont{{Finn}}},
  \bibinfo{author}{\bibfnamefont{S.~D.} \bibnamefont{{Mohanty}}},
  \bibnamefont{and} \bibinfo{author}{\bibfnamefont{J.~D.}
  \bibnamefont{{Romano}}}, \bibinfo{journal}{\prd}
  \textbf{\bibinfo{volume}{60}}, \bibinfo{pages}{121101}
  (\bibinfo{year}{1999}).

\bibitem[{\citenamefont{{Seto}}(2005)}]{seto05}
\bibinfo{author}{\bibfnamefont{N.}~\bibnamefont{{Seto}}},
  \bibinfo{journal}{ArXiv Astrophysics e-prints}  (\bibinfo{year}{2005}),
  \eprint{arXiv:astro-ph/0512212}.

\bibitem[{\citenamefont{{Gustafson} et~al.}(1999)\citenamefont{{Gustafson},
  {Shoemaker}, {Strain}, and {Weiss}}}]{ligonoise}
\bibinfo{author}{\bibfnamefont{E.}~\bibnamefont{{Gustafson}}},
  \bibinfo{author}{\bibfnamefont{D.}~\bibnamefont{{Shoemaker}}},
  \bibinfo{author}{\bibfnamefont{K.}~\bibnamefont{{Strain}}}, \bibnamefont{and}
  \bibinfo{author}{\bibfnamefont{R.}~\bibnamefont{{Weiss}}}
  (\bibinfo{year}{1999}),
  \urlprefix\url{http://www.ligo.caltech.edu/docs/T/T990080-00.pdf}.

\bibitem[{\citenamefont{{Acernese} et~al.}(2005)\citenamefont{{Acernese},
  {Amico}, {Al-Shourbagy}, {Aoudia}, {Avino}, {Babusci}, {Ballardin},
  {Barill{\'e}}, {Barone}, {Barsotti} et~al.}}]{virgo}
\bibinfo{author}{\bibfnamefont{F.}~\bibnamefont{{Acernese}}},
  \bibinfo{author}{\bibfnamefont{P.}~\bibnamefont{{Amico}}},
  \bibinfo{author}{\bibfnamefont{M.}~\bibnamefont{{Al-Shourbagy}}},
  \bibinfo{author}{\bibfnamefont{S.}~\bibnamefont{{Aoudia}}},
  \bibinfo{author}{\bibfnamefont{S.}~\bibnamefont{{Avino}}},
  \bibinfo{author}{\bibfnamefont{D.}~\bibnamefont{{Babusci}}},
  \bibinfo{author}{\bibfnamefont{G.}~\bibnamefont{{Ballardin}}},
  \bibinfo{author}{\bibfnamefont{R.}~\bibnamefont{{Barill{\'e}}}},
  \bibinfo{author}{\bibfnamefont{F.}~\bibnamefont{{Barone}}},
  \bibinfo{author}{\bibfnamefont{L.}~\bibnamefont{{Barsotti}}},
  \bibnamefont{et~al.}, \bibinfo{journal}{Classical and Quantum Gravity}
  \textbf{\bibinfo{volume}{22}}, \bibinfo{pages}{869} (\bibinfo{year}{2005}).

\bibitem[{\citenamefont{{McClelland} et~al.}(1996)\citenamefont{{McClelland},
  {Blair}, and {Sandeman}}}]{aigo}
\bibinfo{author}{\bibfnamefont{D.~E.} \bibnamefont{{McClelland}}},
  \bibinfo{author}{\bibfnamefont{D.~G.} \bibnamefont{{Blair}}},
  \bibnamefont{and} \bibinfo{author}{\bibfnamefont{R.~J.}
  \bibnamefont{{Sandeman}}}, in \emph{\bibinfo{booktitle}{Proceedings of the
  Seventh Marcel Grossman Meeting on recent developments in theoretical and
  experimental general relativity, gravitation, and relativistic field
  theories}} (\bibinfo{year}{1996}), pp. \bibinfo{pages}{1415--+}.

\bibitem[{\citenamefont{{Page} et~al.}(2003)\citenamefont{{Page}, {Nolta},
  {Barnes}, {Bennett}, {Halpern}, {Hinshaw}, {Jarosik}, {Kogut}, {Limon},
  {Meyer} et~al.}}]{page03}
\bibinfo{author}{\bibfnamefont{L.}~\bibnamefont{{Page}}},
  \bibinfo{author}{\bibfnamefont{M.~R.} \bibnamefont{{Nolta}}},
  \bibinfo{author}{\bibfnamefont{C.}~\bibnamefont{{Barnes}}},
  \bibinfo{author}{\bibfnamefont{C.~L.} \bibnamefont{{Bennett}}},
  \bibinfo{author}{\bibfnamefont{M.}~\bibnamefont{{Halpern}}},
  \bibinfo{author}{\bibfnamefont{G.}~\bibnamefont{{Hinshaw}}},
  \bibinfo{author}{\bibfnamefont{N.}~\bibnamefont{{Jarosik}}},
  \bibinfo{author}{\bibfnamefont{A.}~\bibnamefont{{Kogut}}},
  \bibinfo{author}{\bibfnamefont{M.}~\bibnamefont{{Limon}}},
  \bibinfo{author}{\bibfnamefont{S.~S.} \bibnamefont{{Meyer}}},
  \bibnamefont{et~al.}, \bibinfo{journal}{\apjs}
  \textbf{\bibinfo{volume}{148}}, \bibinfo{pages}{233} (\bibinfo{year}{2003}).

\bibitem[{\citenamefont{{Dalal} et~al.}(2003)\citenamefont{{Dalal}, {Holz},
  {Chen}, and {Frieman}}}]{dalal03}
\bibinfo{author}{\bibfnamefont{N.}~\bibnamefont{{Dalal}}},
  \bibinfo{author}{\bibfnamefont{D.~E.} \bibnamefont{{Holz}}},
  \bibinfo{author}{\bibfnamefont{X.}~\bibnamefont{{Chen}}}, \bibnamefont{and}
  \bibinfo{author}{\bibfnamefont{J.~A.} \bibnamefont{{Frieman}}},
  \bibinfo{journal}{\apjl} \textbf{\bibinfo{volume}{585}}, \bibinfo{pages}{L11}
  (\bibinfo{year}{2003}).

\bibitem[{\citenamefont{{Smith} et~al.}(2003)\citenamefont{{Smith}, {Peacock},
  {Jenkins}, {White}, {Frenk}, {Pearce}, {Thomas}, {Efstathiou}, and
  {Couchman}}}]{smith03}
\bibinfo{author}{\bibfnamefont{R.~E.} \bibnamefont{{Smith}}},
  \bibinfo{author}{\bibfnamefont{J.~A.} \bibnamefont{{Peacock}}},
  \bibinfo{author}{\bibfnamefont{A.}~\bibnamefont{{Jenkins}}},
  \bibinfo{author}{\bibfnamefont{S.~D.~M.} \bibnamefont{{White}}},
  \bibinfo{author}{\bibfnamefont{C.~S.} \bibnamefont{{Frenk}}},
  \bibinfo{author}{\bibfnamefont{F.~R.} \bibnamefont{{Pearce}}},
  \bibinfo{author}{\bibfnamefont{P.~A.} \bibnamefont{{Thomas}}},
  \bibinfo{author}{\bibfnamefont{G.}~\bibnamefont{{Efstathiou}}},
  \bibnamefont{and} \bibinfo{author}{\bibfnamefont{H.~M.~P.}
  \bibnamefont{{Couchman}}}, \bibinfo{journal}{\mnras}
  \textbf{\bibinfo{volume}{341}}, \bibinfo{pages}{1311} (\bibinfo{year}{2003}).

\bibitem[{\citenamefont{{Nakar} et~al.}(2005)\citenamefont{{Nakar}, {Gal-Yam},
  and {Fox}}}]{nakar05}
\bibinfo{author}{\bibfnamefont{E.}~\bibnamefont{{Nakar}}},
  \bibinfo{author}{\bibfnamefont{A.}~\bibnamefont{{Gal-Yam}}},
  \bibnamefont{and} \bibinfo{author}{\bibfnamefont{D.~B.} \bibnamefont{{Fox}}},
  \bibinfo{journal}{ArXiv Astrophysics e-prints}  (\bibinfo{year}{2005}),
  \eprint{arXiv:astro-ph/0511254}.

\bibitem[{\citenamefont{{Flanagan} and {Hughes}}(1998)}]{flanagan98}
\bibinfo{author}{\bibfnamefont{{\'E}.~{\'E}.} \bibnamefont{{Flanagan}}}
  \bibnamefont{and} \bibinfo{author}{\bibfnamefont{S.~A.}
  \bibnamefont{{Hughes}}}, \bibinfo{journal}{\prd}
  \textbf{\bibinfo{volume}{57}}, \bibinfo{pages}{4535} (\bibinfo{year}{1998}).

\bibitem[{\citenamefont{{Milosavljevi{\'c}} and {Phinney}}(2005)}]{milos05}
\bibinfo{author}{\bibfnamefont{M.}~\bibnamefont{{Milosavljevi{\'c}}}}
  \bibnamefont{and} \bibinfo{author}{\bibfnamefont{E.~S.}
  \bibnamefont{{Phinney}}}, \bibinfo{journal}{\apjl}
  \textbf{\bibinfo{volume}{622}}, \bibinfo{pages}{L93} (\bibinfo{year}{2005}).

\bibitem[{\citenamefont{{Armitage} and {Natarajan}}(2002)}]{armitage02}
\bibinfo{author}{\bibfnamefont{P.~J.} \bibnamefont{{Armitage}}}
  \bibnamefont{and}
  \bibinfo{author}{\bibfnamefont{P.}~\bibnamefont{{Natarajan}}},
  \bibinfo{journal}{\apjl} \textbf{\bibinfo{volume}{567}}, \bibinfo{pages}{L9}
  (\bibinfo{year}{2002}).

\bibitem[{\citenamefont{{Pen}}(2004)}]{pen04}
\bibinfo{author}{\bibfnamefont{U.-L.} \bibnamefont{{Pen}}},
  \bibinfo{journal}{New Astronomy} \textbf{\bibinfo{volume}{9}},
  \bibinfo{pages}{417} (\bibinfo{year}{2004}).

\bibitem[{\citenamefont{{Seljak} and {Holz}}(1999)}]{seljakholz}
\bibinfo{author}{\bibfnamefont{U.}~\bibnamefont{{Seljak}}} \bibnamefont{and}
  \bibinfo{author}{\bibfnamefont{D.~E.} \bibnamefont{{Holz}}},
  \bibinfo{journal}{\aap} \textbf{\bibinfo{volume}{351}}, \bibinfo{pages}{L10}
  (\bibinfo{year}{1999}).

\bibitem[{\citenamefont{{Sesana} et~al.}(2004)\citenamefont{{Sesana}, {Haardt},
  {Madau}, and {Volonteri}}}]{sesana04}
\bibinfo{author}{\bibfnamefont{A.}~\bibnamefont{{Sesana}}},
  \bibinfo{author}{\bibfnamefont{F.}~\bibnamefont{{Haardt}}},
  \bibinfo{author}{\bibfnamefont{P.}~\bibnamefont{{Madau}}}, \bibnamefont{and}
  \bibinfo{author}{\bibfnamefont{M.}~\bibnamefont{{Volonteri}}},
  \bibinfo{journal}{\apj} \textbf{\bibinfo{volume}{611}}, \bibinfo{pages}{623}
  (\bibinfo{year}{2004}).

\bibitem[{\citenamefont{{Menou} et~al.}(2001)\citenamefont{{Menou}, {Haiman},
  and {Narayanan}}}]{menou01}
\bibinfo{author}{\bibfnamefont{K.}~\bibnamefont{{Menou}}},
  \bibinfo{author}{\bibfnamefont{Z.}~\bibnamefont{{Haiman}}}, \bibnamefont{and}
  \bibinfo{author}{\bibfnamefont{V.~K.} \bibnamefont{{Narayanan}}},
  \bibinfo{journal}{\apj} \textbf{\bibinfo{volume}{558}}, \bibinfo{pages}{535}
  (\bibinfo{year}{2001}).

\bibitem[{\citenamefont{{Haehnelt}}(2003)}]{haehnelt03}
\bibinfo{author}{\bibfnamefont{M.~G.} \bibnamefont{{Haehnelt}}},
  \bibinfo{journal}{Classical and Quantum Gravity}
  \textbf{\bibinfo{volume}{20}}, \bibinfo{pages}{31} (\bibinfo{year}{2003}).

\bibitem[{\citenamefont{{Koushiappas} and {Zentner}}(2005)}]{koushiappas06}
\bibinfo{author}{\bibfnamefont{S.~M.} \bibnamefont{{Koushiappas}}}
  \bibnamefont{and} \bibinfo{author}{\bibfnamefont{A.~R.}
  \bibnamefont{{Zentner}}}, \bibinfo{journal}{Astrophys. J. in press; ArXiv
  Astrophysics e-prints}  (\bibinfo{year}{2005}), \eprint{astro-ph/0503511}.

\end{thebibliography}
\end{document}